\documentclass[prl,twocolumn]{revtex4}%
\usepackage{amsmath}
\usepackage{amssymb}
\usepackage{graphicx}
\usepackage{dcolumn}
\usepackage{bm}
\usepackage{amsmath}
\usepackage{graphicx}
\usepackage{amsfonts}
\usepackage{subfigure}
\usepackage{graphicx}
\usepackage{float}

\usepackage{amssymb}%

\newcommand{\ket}[1]{\ensuremath{\left\vert #1 \right\rangle}}

\begin{document}
\title{Emergence of spatial spin-wave correlations in a cold atomic gas}
\date{\today }
\author{Y. O. Dudin, F. Bariani, and A. Kuzmich}
\affiliation{School of Physics, Georgia Institute of Technology, Atlanta, Georgia
30332-0430}

\begin{abstract} Rydberg spin waves are optically excited in a quasi-one-dimensional atomic sample of Rb atoms.  Pair-wise spin-wave correlations are observed by a spatially selective transfer of the quantum state onto a light field and photoelectric correlation measurements of the light. The correlations are interpreted in terms of the dephasing of multiply-excited spin waves by long-range Rydberg interactions.
\end{abstract}
\maketitle

\begin{figure*}[hbt]
\includegraphics[scale=0.9]{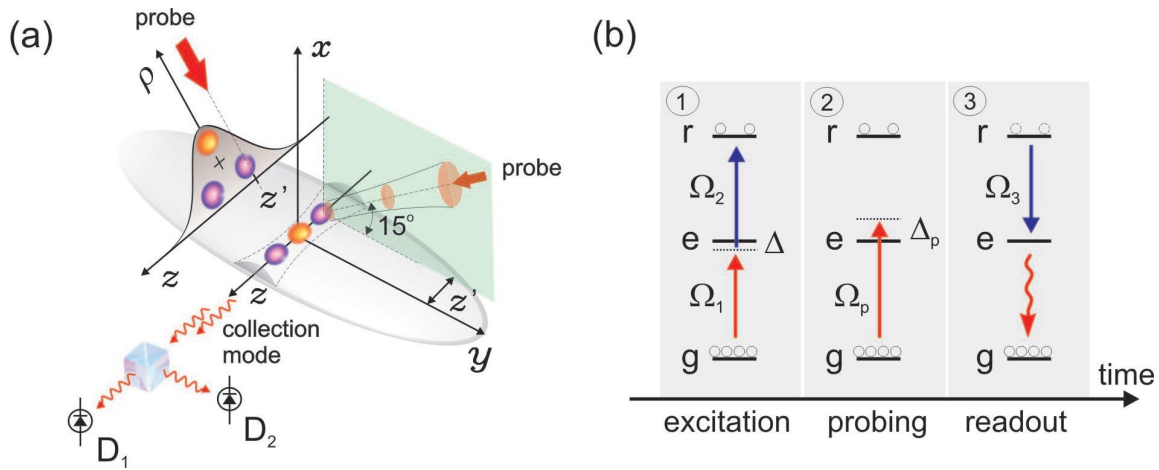}
\caption{(color online). (a) Experimental geometry: an atomic sample of temperature $T\simeq 10$ $\mu$K is produced in an optical lattice formed by a retro-reflected 782 nm laser beam propagating along the y-axis. The quantization axis is defined by a 4.3 G bias magnetic field along the $x$-axis. Localized spin waves are formed in the sample as a result of Rydberg excitation blockade and spin-wave dephasing. The spatial structure is revealed by scanning the focused waist of the probe beam along the $z-$axis of the two-photon excitation region.  Either a single (orange) or double (blue) excitation is formed in an approximately Gaussian density profile $\rho$ of the atomic sample. The probe beam destroys the phase-matched spin waves in a spatially-selective manner, affecting the statistics of the retrieved field. The latter is measured by a Hanbury Brown-Twiss type setup consisting of a beamsplitter and two single-photon detectors D$_1$ and D$_2$. (b) Experimental sequence of spin-wave creation, probing, and readout.}  \label{fig:setup}
\end{figure*}

The realization of model systems of many particles with precisely controlled and widely tunable interactions is of considerable interest to improving our understanding of many-body quantum physics. Cold atomic gases have become a fruitful platform for such studies as they facilitate experiments under well-understood conditions \cite{polkovnikov,bloch}. Collisional interactions based on low-energy s-wave scattering have been used in such studies, with their strength tuned by Feschbach resonances \cite{mandel,anderlini}. Atoms must be cooled to ultra-low temperatures for these short-range interactions to dominate the dynamics.

In contrast, Rydberg-level atoms possess long-range interactions which scale as $1/R^6$ in the van der Waals regime and $1/R^3$ in the dipole-dipole regime. The strength of these interactions for values of the principle quantum number $n\gtrsim 100$ can be comparable to the strength of the Coulomb interaction between ions. This leads to Rydberg excitation blockade, in which a single excited atom can suppress the excitation of multiple nearby atoms \cite{adams,lukin,swm,viteau,singer,vogt,tong,liebisch,heidemann,pritchard}. The multi-atom blockade mechanism has attracted considerable interest for studies at the interface of atomic and condensed matter physics \cite{gorshkov,weimer,breyel,garttner,olmos1,sela,ates}.

The interactions between atoms are switched on and off using laser fields tuned to the frequency between the Rydberg level and the ground level. In an ensemble of atoms, such coherent driving generates collective Rydberg spin waves. Spin waves can be mapped onto light using quantum state transfer techniques. Subsequent photoelectric detection of the light permits one to extract their properties \cite{fleischhauer,matsukevich,bariani,barianilong,stanojevic}.
Crystals composed of many such spin waves can be generated dynamically by an adiabatic sweep of laser detuning through the so-called Rydberg staircase, with the number of excited atoms determined by the sweep parameters \cite{pohl,schachenmayer}. Recently, Ates and Lesanovsky have shown that an atomic gas undergoing continuous-wave, on-resonance Rydberg excitation can develop long-range correlations arising from the interplay of atomic interactions and dynamic spin-wave generation \cite{ates}.

Here we report observations of interaction-induced localization of Rydberg spin waves optically-excited in an ultra-cold atomic gas. The experimental apparatus and methods have been described in Ref. \cite{dudin2012}, where we produced single Rydberg spin-wave excitations generated in a $\sim 15$ $\mu$m-long ensemble. In this work we employ similar techniques to study pair-wise spin-wave correlations in a longer, up to $\sim 100$ $\mu$m, ensemble.

A sample of $^{87}$Rb atoms of peak density $\rho_0 \simeq 10^{12}$ cm$^{-3}$ is prepared in an optical lattice. The lattice is shut off, and an ensemble is driven in two-photon resonance between the ground $|g\rangle=|5s_{1/2}\rangle$ and a Rydberg $|r\rangle=|102s_{1/2}\rangle$ levels with a laser excitation pulse of duration $\tau=200$ ns, Fig. 1.
The transverse size (Gaussian waists $w_x \approx w_y \simeq 6$ $\mu$m) of the Rydberg excitation region is determined by
the overlap of the two-photon excitation laser fields $\Omega_1$ at 795 nm and $\Omega_2$ at 475 nm. The longitudinal size
(waist $w_z$ variable  between 10 and 60 $\mu$m) is determined by the sample size along $z$. The number of atoms $N$ involved in the excitation is therefore $(1-5)\times 10^3$.
The values of $\Omega_1$ and $\Omega_2$ are chosen such that the multi-atom pulse excitation area $\theta\equiv \sqrt{N}\Omega \tau \sim 1$, where
$\Omega = \Omega_1 \Omega_2 / \Delta$ is the single-atom, two-photon Rabi frequency, and $\Delta$ is the laser detuning from the intermediate level $|e\rangle=|5p_{1/2}\rangle$. Since $w_x$ and $w_y$ are smaller than the effective atomic interaction range, quasi-1D spin waves can be realized for $w_z > w_x, w_y$. In the limit of instantaneous excitation
($\tau \rightarrow 0$), the spin-wave can be described by a coherent state $\ket{\Psi_0} =  \sum_{\alpha = 0}^{N} (c_{\alpha}/\sqrt{\alpha!}) S^{\dagger \alpha}_{\mathbf{k}_0} \ket{G}$, where the creation operator for the collective excitations is $S^{\dagger}_{\mathbf{k}_0} = (1/\sqrt{N}) \sum_{\mu} e^{- i \mathbf{k}_0 \cdot \mathbf{r}_{\mu}} \hat{\sigma}_{\mu}^{rg} $ and $\hat{\sigma}_{\mu}^{rg}$ is the raising operator for the atom $\mu = [1,N]$, and $|G\rangle \equiv \prod^N_{i=1}|g\rangle_i$ is the collective ground state \cite{bariani}. Here ${\bf k}_0\equiv{\bf k}_{1} + {\bf k}_{2}$, and
${\bf k}_{1,2}$ are wave-vectors of the excitation laser fields $\Omega_{1,2}$. In our measurement the intensity of the excitation is chosen so as to create on average approximately one excitation in the sample \cite{bariani,dudin2012}. Correspondingly a Poissonian distribution of unit mean, $c_{\alpha}=1/\sqrt{e\alpha !}$, is used for the initial spin-wave excitations in the numerical simulations. For finite values of $\tau$, Rydberg-level interactions suppress the excitation of nearby atoms, resulting in a reduction of amplitudes $c_{\alpha}$ for $\alpha \geq 2$.

 The collective multi-atom state evolves during the ensuing storage period of duration $T_s$. For $T_s$ shorter than the motional dephasing time the spin-wave dynamics are dominated by the interactions of Rydberg atoms. As a result, the coherently excited sample evolves into a set of localized spin waves, each one containing no more than a single excitation. This may be viewed as spin-wave antibunching, in which the probability of having excitations in two different regions of the sample is suppressed for nearby regions.

 \begin{figure}[t]
\includegraphics[scale=0.5]{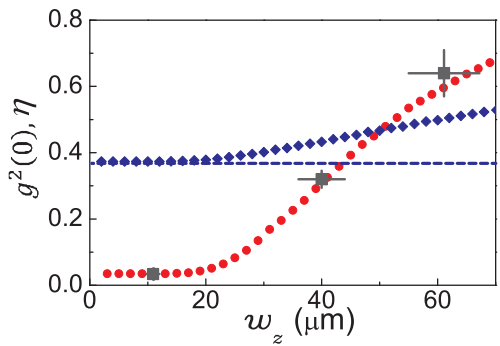}
\caption{(color online). Second-order intensity correlation function $g^{(2)}(0)$ and the average number of spin-wave excitations $\eta$ as a function of the sample length. The measured values of $g^{(2)}(0)$ are shown as squares, whereas the circles and diamonds show calculated $g^{(2)}(0)$ and $\eta$, respectively. In the numerical simulations the interval between the excitation and the retrieval $T_s=0.8$ $\mu$s and the transverse sample waists $w'_{x,y}$ =6.4 $\mu$m are chosen to approximate the conditions of the experiment. The dashed line at $1/e$ shows $\eta$ when all spin waves with more than one excitation have dephased \cite{bariani}.}
\label{fig:g2}
\end{figure}
The spin wave is converted into a retrieved light field by a 1 $\mu$s long read-out field $\Omega_3$ at 475 nm, in resonance with the $|102s\rangle \leftrightarrow |5p_{1/2}\rangle$ transition.  The retrieved field is coupled into a single mode fiber followed by a beam splitter and a pair of single-photon detectors D$_1$ and D$_2$. The photoelectric events on the detectors are recorded
 within gated time intervals determined by the length of the retrieved pulse. Photoelectric detection probabilities for D$_1$ and D$_2$ are calculated as $p_{1,2}=N_{1,2}/N_0$, where $N_{1,2}$ are numbers of recorded events and $N_0$ is the number of experimental trials. Photoelectric detection probability for double coincidences  $p_{12}$ is calculated as $N_{12}/N_0$, where $N_{12}$ is a total number of simultaneous clicks on both detectors.
The second order intensity correlation function at zero delay is given by $g^2(0) = p_{12}/(p_1p_2)$. Using weak classical light pulses within the experimental sequence we have measured $g^2(0)=0.99(2)$, consistent with the expected unity value.

For a sufficiently small sample ($w_z\lesssim 20$ $\mu$m in Fig. 2), the combination of the excitation blockade and interaction-induced dephasing can completely suppress amplitudes for two or more excitations in the retrievable spin wave, realizing a source of single photons for which $g^{(2)}(0) \rightarrow 0$ \cite{dudin2012}. The atoms perform the role of a quantum filter: after the (weak) coherent state of light is mapped onto atoms,  Rydberg interactions dephase the components of the spin wave that contain more than one excitation. The phase-matching-based read-out protocol therefore generates a retrieved field containing the vacuum and the single-photon components only, for which the average number of spin-wave excitations $\eta=1/e$ \cite{bariani}.

 \begin{figure}[t]
\includegraphics[scale=0.42]{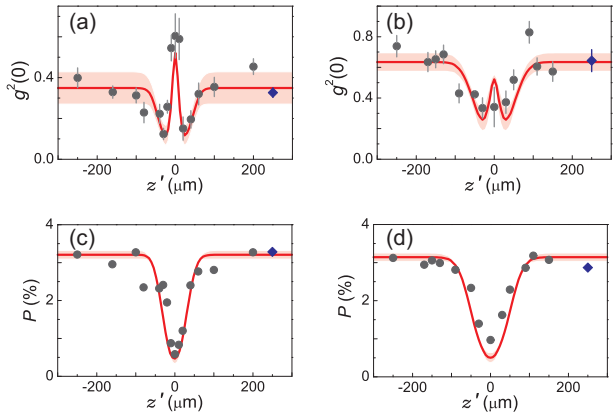}
\caption{(color online). $g^{(2)}(0)$ (top panels) and $P$ (bottom panels).  Left panels are for $w_z=42$ $\mu$m, right panels for $w_z=64$ $\mu$m, with these values extracted by the fluorescence imaging of the atomic cloud. The solid curves are based on computer simulations with $w'_z=42$ $\mu$m for left panels, and $w'_z=64$ $\mu$m for right panels. The shaded regions are for a range of $w'_z$ of 38 to 46 $\mu$m, and 58 to 70 $\mu$m for the left and right panels, respectively, while the results are only weakly sensitive to the variation of $w'_x$ and $w'_y$.}
\label{fig:g2a}
\end{figure}

For a longer ensemble, the $1/R^6$ scaling of the interaction strength in the van der Waals regime can result in well-defined ``interaction volumes" within the sample. In this case, the emergence of a spatial pattern of retrievable spin-waves is expected. In Fig. 2 we display the measured values of $g^{(2)}(0)$ as a function of the ensemble length, together with the results of the numerical simulations of $g^{(2)}(0)$ and of the average number of retrievable spin-wave excitations  $\eta\equiv\langle \hat{S}_{{\bf k}_0}^{\dagger} \hat{S}_{{\bf k}_0} \rangle$ for $T_s=0.8$ $\mu$s. The latter should be compared to its initial value of $1$ for the Poissonian distribution of unit mean.  Both $g^{(2)}(0)$ and $\eta$ rise with sample length due to the increasing survival probability of double excitations.

To look into the spatial structure of the spin wave, we illuminate it with a near-resonant 795 nm probe laser field $\Omega_p$ for 150 ns during the 0.5 $\mu$s time interval between the Rydberg excitation and readout stages, Fig. 1b. The beam, focused down to a $\approx 10$ $\mu$m waist ($1/e^2$ intensity) at the position of the sample, produces a local dephasing of the spin wave via the associated inhomogeneous light shifts and spontaneous emission. The retrieved signal in this case corresponds
to the spin-wave content of the part of the sample unaffected by the probe beam. By scanning the position of the probe beam $z'$, it is, therefore, possible to reveal spatial spin-wave correlations.

The measured $g^{(2)}(0)$ and the probability of photoelectric detection $P\equiv p_1+p_2$ are shown in Fig. \ref{fig:g2a}.
When the beam is sufficiently far from the sample so as not to affect it ($|z'| \geq 0.2$ mm), the measured values are consistent with those for no probe beam (the blue diamond in each panel). As the beam position approaches the sample notable decreases in $g^{(2)}(0)$ and efficiency are observed.
$P$ reaches its minimum value with the beam illuminating the center of the ensemble, while $g^{(2)}(0)$ has a peak whose prominence depends on the length of the cloud.

%Qualitatively these features can be understood as follows: the single excitation is peaked at the center of the ensemble because of the higher atom density, %while the the double excitations are localized at the ends of the ensemble because of the interaction-induced dephasing. When one of the ends of the ensemble is %illuminated by the probe beam, only a single excitation is retrieved with efficiency reduced by the reduced number of atoms participating in the read-out %process. When the probe is focused into the center of the ensemble, the efficiency is further reduced due to a maximum number of atoms removed, while the %correlation function rises because the excitations localized at the ends are untouched.

Qualitatively these features can be understood as follows: the single excitation is peaked at the center of the ensemble because of the higher atom density, while the double excitations are localized at the ends of the ensemble because of the interaction-induced dephasing. The two-photon detection events correspond to two spatially-localized singly-excited spin waves on the two ends of the sample. When the probe beam is focused on one of the ends, the corresponding spin wave is destroyed, thus substantially reducing the probability of a two-photon event $p_{12}$. The reduction of the numerator results in a low value of $g^2(0)$.
When the probe beam illuminates the center of the ensemble, $p_{12}$ is not significantly affected because the corresponding excitations are localized largely at the sample ends. The probabilities $p_{1}$ and $p_2$ are however reduced due to the removal of the singly-excited spin waves at the center of the ensemble. The concomitant reduction of the denominator of $g^2(0)$ leads to an increase of the latter. For the longer ensemble (Fig. \ref{fig:g2a}(b,d)), the center peak is not as pronounced in the simulation because of a higher average number of excitations, and is not evident beyond the statistical uncertainties of the data.

Our numerical modeling employs Monte-Carlo simulations for 100 atoms randomly sampled according to a gaussian density with transverse waists $w'_x = w'_y$ and longitudinal waist $w'_z$, see Ref. \cite{bariani}. A single channel model is used to calculate the interaction-induced detuning \cite{swm}.
The annihilation $\hat{a}_{{\bf k}_0}$ and creation $\hat{a}_{{\bf k}_0}^{\dagger}$ operators of the retrieved field determine $g^{(2)}(0) \equiv \langle\hat{a}_{{\bf k}_0}^{\dagger}\hat{a}_{{\bf k}_0}^{\dagger}\hat{a}_{{\bf k}_0}\hat{a}_{{\bf k}_0} \rangle/\langle\hat{a}_{{\bf k}_0}^{\dagger}\hat{a}_{{\bf k}_0} \rangle^2$.
The retrieval process is described by the linear relation $\hat{a}_{{\bf k}_0} = \sqrt{\epsilon} \hat{S}_{{\bf k}_0} + \sqrt{1 - \epsilon} \hat{\xi}$, where $\epsilon$ is the overall efficiency for the atom-light mapping, field transmission, and photoelectric detection.
If the corresponding operator $\hat{\xi}$ is assumed to be in the vacuum state,
the value of $g^{(2)}(0)$ is independent of linear losses and is equal to the corresponding spin-wave correlation function
$g_s^{(2)}(0) = \langle \hat{S}_{{\bf k}_0}^{\dagger}\hat{S}_{{\bf k}_0}^{\dagger}\hat{S}_{{\bf k}_0}\hat{S}_{{\bf k}_0}\rangle/\langle \hat{S}_{{\bf k}_0}^{\dagger}\hat{S}_{{\bf k}_0}\rangle^2$. In the limit $P\ll1$ the photoelectric detection probability $P\approx \epsilon \eta$.

 The action of the probe beam is simulated by transferring the atoms located in the illuminated part of the ensemble into a state that is not retrieved by the read pulse, and the spin-wave correlation function becomes spatially dependent. The calculated solid curves and shaded regions in Fig. 3 are in in good agreement with the experimental data, which is achieved even though our simulations do not account for the dynamical effects of the excitation blockade, suggesting that the dephasing of Rydberg spin waves is largely responsible for the observed features.

 In practice background light and detector dark counts contribute to the the photoelectric detection probabilities, so that the measured value $g^{(2)}(0)$ will differ from the quantity $g_s^{(2)}(0)$. To study the role of background photoelectric detection events in our experiment, we vary the probability of a photoelectric detection event $P$ by inserting  optical filters into the retrieved field with no probe beam used, Fig. 4(a). The data shows that over the dynamic range of our experiment the measured values of $g^{(2)}(0)$ are not significantly affected by losses. To further assess the influence of the background counts, in Fig. 4(b) we display both $g^{(2)}(0)$ and the inferred quantity $g^{(2)}_s(0)$. The latter values are somewhat lower when the probe beam is near the center of the sample, with the same functional dependence.
\begin{figure}[tb]
\includegraphics[scale=0.6]{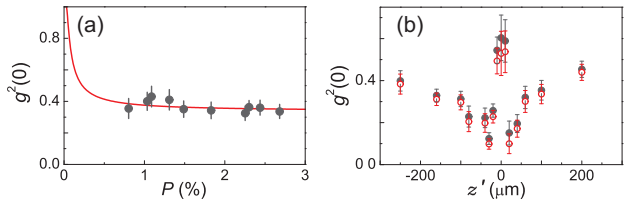}
\caption{(color online). (a) $g^{(2)}(0)$ as a function of photoelectric detection probability $P$ that is varied by inserting partially transmitting optical filters into the retrieved field mode. The solid line is the function $1+(g^{(2)}_s(0)-1)(1-\frac{B}{P})^2$ derived by choosing the background mode described by an annihilation operator $\hat{\xi}$ to be in a coherent state with average photon number $B$. Here $P$ is a sum of contribution $P_0$ due to the retrieved field and of background contribution $B$. (b) $g^{(2)}(0)$ as a function of the probe beam position along the atomic sample $z'$. Filled circles are for the case without subtraction of background counts as in Fig. 3 and open circles are for the inferred values of $g^{(2)}(0)$ that would have been measured instead if there were no background counts.}
\label{gsi_graph}
\end{figure}

 \begin{figure}[bt]
\includegraphics[scale=0.86]{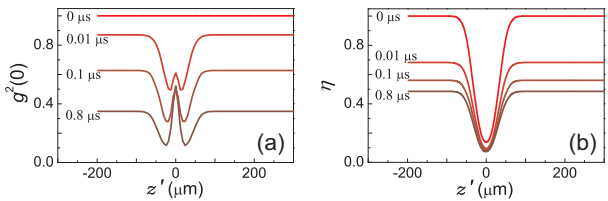}
\caption{(color online). (a) Intensity correlation function at zero delay, $g^{(2)}(0)$. (b) intensity of the signal, $\eta$, both as a function of the probe beam position along the z-axis, for storage periods $T_s$ up to 0.8 $\mu$s, with $w'_{x,y}$ = 8 $\mu$m, and $w'_{z}$ =42 $\mu$m.}
\label{fig:g2}
\end{figure}

In the future it may be possible to observe the time evolution of the spin waves by achieving longer coherence times or a higher temporal resolution of the excitation-probe-retrieval sequence. As an example, we show $g^2(0)$ and $\eta$ normalized by its value when no probe beam is employed in Fig. 5 (a) and (b), respectively, for varying values of spin-wave storage time $T_s$. Both $g^2(0)$ and $\eta$ decrease with $T_s$ reflecting the growth of the Rydberg interaction volume. The peak of $g^2(0)$ at $z'\approx 0$ becomes prominent at $T_s\simeq 0.8$ $\mu$s when the spin-waves with more than two excitations have mostly dephased.

We have demonstrated the emergence of localized collective atomic excitations in a strongly interacting atomic gas. In the future the extent of atomic correlations may be increased by using a smaller atomic sample and employing lasers with narrower linewidths. It is likely that the spin-wave coherence is considerably more sensitive than a single-atom coherence to various broadening mechanisms. Therefore imaging techniques of the type described in Refs. \cite{olmos2,gunter} may be advantageous for observing Rydberg crystals.

We thank P. Goldbart, B. Kennedy, M. Pustilnik, and A. Zangwill for discussions. This work was supported by the Atomic Physics Program and the Quantum Memories MURI of the AFOSR and the NSF.

\end{document}